\documentclass{elsarticle}

\usepackage{lineno,hyperref}
\modulolinenumbers[5]
\usepackage{amsmath}
\usepackage[]{algorithm2e}
\makeatletter
\renewcommand{\@algocf@capt@plain}{above}
\makeatother

\journal{Journal of \LaTeX\ Templates}

\begin{document}

\begin{frontmatter}

\title{Spectrum Sensing and Resource Allocation for 5G Heterogeneous Cloud Radio Access Networks}


\author[mymainaddress1]{Hossein Safi}
\author[mymainaddress2]{A. M Montazeri}
\author[mymainaddress3]{Javane Rostampoor}
\author[mymainaddress3]{Saeedeh Parsaeefard\corref{mycorrespondingauthor}}
\cortext[mycorrespondingauthor]{Corresponding author}
\ead{saeede.parsaeefard@gmail.com}

\address[mymainaddress1]{Faculty of Electrical Engineering, Shahid Beheshti University, Tehran, Iran}
\address[mymainaddress2]{Communications Technology (CT) Research Faculty, ICT Research Institute,Tehran, Iran}
\address[mymainaddress3]{Department of Electrical and Computer Engineering, The
University of Toronto, Toronto, Canada}

\begin{abstract}
In this paper, the problem of opportunistic spectrum sharing for the next generation of wireless systems empowered by the cloud radio access network (C-RAN) is studied. More precisely, low-priority users employ cooperative spectrum sensing to detect a vacant portion of the spectrum that is not currently used by high-priority users. The design of the scheme is to maximize the overall throughput of the low-priority users while guaranteeing the quality of service of the high-priority users. This objective is attained by optimally adjusting spectrum sensing time with respect to imposed target probabilities of detection and false alarm as well as dynamically allocating and assigning C-RAN resources, i.e., transmit powers, sub-carriers, remote radio heads  (RRHs), and base-band units. The presented optimization problem is non-convex and NP-hard that is extremely hard to tackle directly. To solve the problem, a low-complex iterative approach is proposed in which sensing time, user association parameters and transmit powers of RRHs are alternatively assigned and optimized at every step. Numerical results are then provided to demonstrate the necessity of performing sensing time adjustment in such systems as well as balancing the sensing-throughput tradeoff.
\end{abstract}


\end{frontmatter}

\linenumbers
\begin{nolinenumbers}
	\section{Introduction}
The fifth generation (5G) of wireless networks leverages emerging and promising technologies to revolutionize our lives by offering remarkable services e.g., massive IoT  \cite{andrews2014will,shafi20175g,bangerter2014networks}. Since 5G provides a variety of different services, the main concern is how to efficiently allocate and manage the  limited network resources for these emerging data-hungry services. To confront these challenges, far more sensible network designs and the introduction of new strategies for enhancing spectral and energy efficiencies are required.
	
	In this regard, cloud radio access network (C-RAN) is considered as a potential enabler to allow the realization of the vision of 5G, i.e., very dense deployments and centralized processing \cite{peng2014heterogeneous,rost2014cloud}. Compared to the traditional RAN,  via centralizing powerful processing resources i.e., baseband units (BBUs), C-RAN brings more spectral and energy efficiency that leads to a drastic reduction in investment and maintenance costs\cite{checko2015cloud}. Also, to improve
	the flexibility of network resource allocation, physical resources of the network can be virtually shared  between different operators where the concept of virtual wireless networks (VWNs), also known as network slicing, has been emerged in this context \cite{richart2016resource}. By employing network slicing, in a same physical network, radio resources can be dynamically allocated to  different logical network slices based on the different quality of services (QoS) demands, hence 5G networks will be tailored to meet users' specific QoS requests. Furthermore, extended
	dynamic spectrum allocation (eDSA) has been proposed to facilitate flexible and efficient utilization of the available radio resources from different spectrum bands in heterogeneous networks (HetNets) \cite{caso2017toward,yang2016advanced,5g2016view}. Therefore, a new horizon for the effective employment of spectrum management techniques is opened such as spectrum aggregation, spectrum pooling, mutual spectrum renting, and licensed-assisted access (LAA) \cite{bhushan2014network,kliks2015spectrum,bogucka2015dynamic,monserrat2015metis}.

However, in a heterogeneous environment, network designers may face formidable challenges to fulfill certain traffic demands and various licensing schemes due to strict requirements for protecting high-priority users. More specifically, in a heterogeneous VWN with a shared spectrum access policy,  high-priority VWN (HVWN) users are able to exclusively access the allocated spectrum while low-priority VWN (LVWN) users have to sense the spectrum and transmit in an opportunistic way whenever the HVWN users are in the idle mode and consequently spectrum is vacant. Hence, spectrum sensing has a pivotal role for the effective implementation of dynamic spectrum sharing scenarios.

 There are two key parameters for performing spectrum sensing by LVWN users: 1) the probability of correctly detecting active HVWN users; 2) the probability of false alarm, i.e.,  the probability of misinterpreting the noise as HVWN user's signal while the spectrum is actually vacant. From the HVWN user's perspective, the higher the probability of detection, the less radio interference it receives. On the other hand, from the LVWN user's perspective, the lower the false alarm, the more opportunity for the user to utilize the vacant spectrum and transmit its data. Clearly, there exists a reconciliation between priorities in determining the values of these parameters. Therefore,  for an efficient spectrum sensing  algorithm, the probability of detection should be close to one while the probability of false alarm
 should be decremented in the most possible extend \cite{liang2008sensing}. Since these parameters are directly related to the sensing time, holding these probabilities in the acceptable criteria induces a sensing-throughput tradeoff where, to meet target values of detection and false alarm probabilities and to maximize the achievable throughput of the LVWN, the sensing time should be incremented and decremented, respectively. Consequently, it is of essential to consider sensing time as a new optimization variable subject to the target values of sensing parameters and other system limitations.

From the system-level's perspective, sensing strategies can be classified into two categories, i.e., cooperative spectrum sensing and non-cooperative spectrum sensing. In the cooperative spectrum sensing method, multiple sensing nodes cooperatively sense the spectrum and report their sensing results to a fusion center to decide on vacant channel. In the non-cooperative method, each sensing node itself makes a decision on the vacant channel without any cooperation. It is evident that by employing cooperative spectrum sensing method, the probability of detection can be greatly increased which leads to a better resource management scheme in comparison with non-cooperative method \cite{akyildiz2011cooperative}. Indeed, C-RAN can be considered as a potential enabler for the cooperative spectrum sensing strategies and facilitates their implementation by using BBU pool as a powerful and centralized fusion center. Obviously, a centralized processing unit is capable of integrating sensing and opportunistic access to the the spectrum, and hence, can mitigate interference in the maximum extent.

After performing spectrum sensing, in the transmission phase, efficient resource allocation schemes should be exploited to achieve high throughput for the LVWN users. On this basis and motivated by the sensing-throughput tradeoff, in our study, we adopt the concept of spectrum sensing and attempt to jointly optimize sensing time and transmission parameters in an opportunistic-based spectrum-sharing system which is empowered by network slicing and C-RAN.
\subsection{Realated Works}	
This work can be placed at the intersection of two research areas: 1) dynamic resource allocation in VWNs, 2) cooperative spectrum sensing in 5G networks. Radio resource allocation in C-RAN empowered networks has been paid many attentions in the literature \cite{parsaeefard2018user,huang2017resource,fang2016energy,peng2015energy,parsaeefard2017dynamic}. For instance, in \cite{parsaeefard2018user} user association in C-RAN with massive MIMO configuration is studied to  achieve high rates for cell-edge users. By employing a game-theoretic mechanism with incomplete information, authors in \cite{huang2017resource} proposed a framework for  resource allocation problem, where the device-to-device (D2D) links utilize common resources of multiple cells and each player's transmission parameter (information) is unknown to other players. Also, in \cite{fang2016energy}, the problem of optimizing subchannel assignment and power allocation to maximize the energy efficiency for the downlink non-orthogonal multiple access (NOMA) network is investigated. Reference \cite{peng2015energy} dealt with a joint optimization solution for resource block
assignment and power allocation to maximize energy efficiency performances in the orthogonal frequency-division multiple access
(OFDMA)-based CRAN. Authors in \cite{parsaeefard2017dynamic} studied dynamic resource allocation for virtualized wireless
networks in Massive-MIMO-aided and fronthaul-limited C-RAN. Moreover, in a comprehensive literature review of \cite{related0}, some of the works that have already been done to achieve wireless network virtualization are provided and important aspects of wireless network virtualization are also presented.

Due to the importance of spectral efficiency in 5G and its potential solution, eDSA, a myriad of works are dedicated to study spectrum sharing and management in 5G networks \cite{kliks2015spectrum,gao2016cooperative,lin2014spectrum}. Reference \cite{kliks2015spectrum}  addressed the problem of flexible spectrum sharing by the application of adaptive licensing among interested stakeholders. Also in \cite{gao2016cooperative}, a novel best cooperative mechanism for wireless energy harvesting and spectrum sharing in 5G networks is proposed. Accordingly, in this scheme, data transfer and energy harvesting are finished in the designed timeslot mode while secondary users harvest energy from both ambient signals and primary user's signals. Moreover, authors in \cite{lin2014spectrum} considered two main scenarios of implementing cognitive radio systems, i.e., underlay and overlay scenarios and studied how D2D users should access spectrum in such scenarios. Meanwhile,  the problem of wireless resource virtualization with D2D communication underlaying the LTE network is investigated in \cite{moubayed2015wireless}.
	
Cooperative spectrum sensing has also been widely regarded in the literature \cite{cichon2016energy,khanikar2018cooperative,sardellitti2013joint,related1,related2,xu2018joint}. In \cite {cichon2016energy}, a novel decentralized cooperative spectrum sensing scheme capable of operating in the absence of dedicated reporting channels is presented to offer the freedom of using multiple bits for transmitting a quantized version of the sensing test statistics between secondary user nodes.  In \cite{khanikar2018cooperative}, the authors studied spectrum-sharing between a primary licensee and a group of secondary users. Indeed, in order to enable access to unused licensed spectrum, a secondary user has to monitor licensed bands and opportunistically transmit whenever no primary signal is detected. Moreover in \cite{sardellitti2013joint}, a joint optimization of detection thresholds, sensing time and power allocation over a multichannel transmission scheme, incorporating the loss due to running a fully distributed cooperative sensing algorithm is proposed. To improve the sensing performance, ref. \cite{related1} proposed cooperative spectrum sensing to combat shadowing/fading effects which makes distinguishing between deep fade and vacant band difficult. In \cite{related2}, to enhance the capacity of unlicensed users, the problem of capacity maximization is derived based on optimizing cooperative spectrum sensing parameters which is a function of the number of cooperative users and control channel bandwidth. In \cite{xu2018joint}, a capacity maximization problem of half-duplex (HD) and full-duplex (FD) cognitive OFDM-NOMA systems is considered where an iterative framework is proposed to jointly optimize the sensing time, the user assignment on each sub-carrier, and the allocated power for each user. However, instead of considering the induced interference in each sub-carrier,  the spectrum sensing constraint of \cite{xu2018joint} is only considered to restrict a sum of total interference over the whole part of the spectrum band of the primary network which is less practical, and also its system model does not include the concept of C-RAN as well as network slicing.
	
To the best of our knowledge, non of the previous works considers the problem of joint spectrum sensing time optimization and dynamic resource allocation in a slice-based 5G system with considering C-RAN constraints. To fill this gap, we present an opportunistic spectrum sharing approach for a cloud-based 5G system with network slicing, and investigate the problem of joint spectrum sensing time optimization and dynamic resource allocation where C-RAN and slicing constraints are taken into account.
	
	\subsection{Contributions and Paper Structure}
	In this paper, the problem of dynamic resource allocation  in a C-RAN empowered slice-based 5G network is considered. For this network that includes users with different priorities, i.e., LVWN users and HVWN users, we attempt to maximize the overall throughput of the LVWN users while guaranteeing the QoS of the HVWN users\footnote{Note that, in this paper, when an HVWN user is active, its QoS will be guranteed by avoiding genreting radio interference from LVWN users through imposing high probabilitby of detection for the spectrum sensing scheme.}   via joint sensing time optimization and optimum resource allocation in terms of the transmit powers of the LVWN users,  sub-carriers, and user association parameters. To hold the QoS of the HVWN users and also have a high chance of successful use of the available frequency bands for the LVWN users, we impose target probabilities of detection and false alarm for spectrum sensing as constraints. Then the problem of joint sensing time optimization, user assignment and resource allocation is formulated to achieve overall throughput maximization of the LVWN users.
	
	Our introduced optimization problem is non-convex and NP-hard  that is extremely hard to tackle directly. The original problem is then solved in three steps which are optimized in alternation: 1) sensing time optimization subproblem that is converted to a convex optimization problem by employing some mathematical simplifications; 2) user association subproblem, and
	by converting its constraints into a linear form, this problem leads to an integer linear programming (ILP) and can be optimally solved by using internal MOSEK solver of the CVX package \cite{mosek}; 3) power allocation subproblem, and after employing successive convex approximation (SCA), it is  converted into a relaxed convex optimization problem.
	
	The rest of this paper is organized as follows. Section II represents our system model and problem formulation. The proposed solutions and corresponding algorithms are introduced in Section III, followed by computational complexity and convergence analysis in Section IV. Simulation results are provided in Section V, and finally, the paper is concluded in Section VI.
	\section{Network Structure and Problem Formulation}
We assume that the regional coverage for the LVWN is provided by a C-RAN structure as shown in Fig. 1. A set of Slices, $\mathcal{S} = \{1,\cdots,S\}$,  are served by a set of remote radio heads (RRHs) $\mathcal{R} = \{1,\cdots,R\}$. Supported by limited-capacity fronthaul links, RRHs are connected to a set of  $\mathcal{B} = \{1,\cdots,B\}$ BBUs to process the baseband signals. Each slice $ s \in \mathcal{S} $ has a set of $\mathcal{N}_s = \{1,\cdots,N_s\}$ users and
	requires a minimum reserved rate $\Re_s^{\text{rsv}}$ to support variable data demands of its users. To protect the users of the HVWN from interference, by employing cooperative spectrum sensing, RRHs can opportunistically access the idle spectrum band of the HVWN that is divided into a set of sub-carriers $\mathcal{K} = \{1,\cdots,K\}$. 
	\begin{figure}[!t]
		\centering
		\includegraphics[width=3.45in]{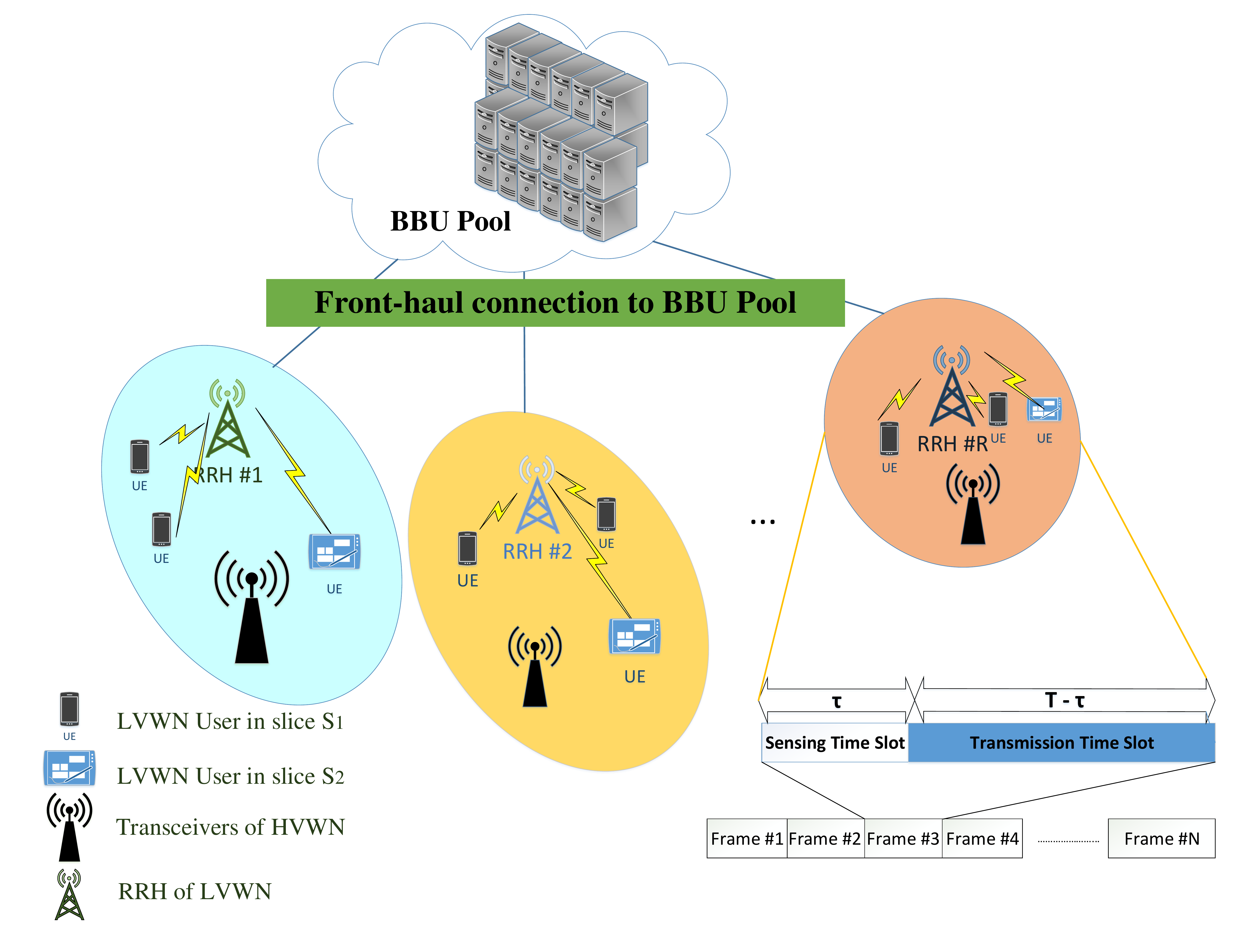}
		\caption{System model and the structure of sensing-transmission time frames.}
		\label{fig1}
	\end{figure}

	Downlink transmission for the set of RRHs is considered where the transmission time is divided into frames with a fixed length $T$, as depicted in Fig. 1. More specifically, each frame consists of three time slots, i.e., sensing time slot, reporting and decision time slot, and transmission time slot are incorporated in each frame. The duration of the sensing time slot for each sub-carrier and each RRH is denoted by $\tau _{r,k}$, which is a variable to be optimized to attain maximum achievable throughput of the LVWN users subject to the protection of the HVWN users. In the reporting and decision time slot, each RRH sends its measurement to a \emph{fusion center} located in C-RAN and decision can be made through data fusion \cite{liang2008sensing}. Obviously, it can be seen that the reporting and decision time slot adds a fixed-time overhead and without loss of generality it will be ignored in the sequel. Next, the optimization problem considered in this work will be described in details. Meanwhile, the summary of employed notations in this work is provided in Table \ref{tabs}.
	\begin{table}
		\def\tablename{Table}
		\caption{Summary of Employed Notations} 
		\centering 
		\begin{tabular}{l l} 
			\hline\hline \\[-1.2ex]
			{\bf Parameter} & {\bf Definition} \\ [.5ex] 
			\hline\hline \\[-1.5ex]
			$\mathcal{S}$          &  Set of slices
			\\[.1ex]
			$\mathcal{R}$          &  Set of RRHs
			\\[.1ex]
			$\mathcal{B}$          &  Set of BBUs
			\\[.1ex]
			$\mathcal{N}_s$         &  Set of users at each slice $s$
			\\[.2ex]
			$\mathcal{K}$          &  Set of sub-carriers
			\\[.2ex]
			$\Re_s^{\text{rsv}}$          &   Minimum reserved rate at each slice $s$
			\\[.2ex]
			$T$          &  Length of transmission time
			\\[.2ex]
			$\tau_{r,k}$          &  Duration of the sensing time slot
			\\[.2ex]
						${\bar P^k}$     & Target probability of detection at sub-carrier $k$
			\\[.2ex]
			${\bar P_{fa}^k}$     & Target probability of false alarm at sub-carrier $k$
			\\[.2ex]
			${P_d^k}$              & Detection probability at sub-carrier $k$
			\\[.3ex]
			$P^1$          & The probability of a HVWN being active
 			\\[.2ex]
			${h_{r,k}^{HU}}$              & The channel coefficient between RRH $r$ and the HVWN user at sub-carrier $k$
			\\[.2ex]
			$h_{r,k,{n_s}}$         &  Channel power gain of the link between RRH $r$ and user $n_s$ at sub-carrier $k$
			\\[.2ex]
			$\gamma_p$            & Received SNR of the HVWN user measured at the assigned RRH
			\\[.2ex]
			$\gamma_{b,k,{n_g}}^0$  & The SINR of the LVWN user $n_s$ (in the absence of HVWN user)
			\\[.2ex]
			$\gamma_{b,k,{n_g}}^1$  &  The SINR of the LVWN user $n_s$ (in the presence of HVWN user)
			\\[.2ex]
			$I_{r,k,{n_s}}$  & Induced interference to user $n_s$ from the other LVWN users
			\\[.2ex]
			$\nu_{sa}$          & Sampling frequency
			\\[.2ex]
			$O^\text{max}$    & Maximum number of users that each BBU can support
			\\[.2ex]
			$C_{r,b}^{\text{max} }$ &Maximum capacity of
			front-haul link between
			\\[.2ex]
			&  RRH $r$ and BBU $b$
			\\[.2ex]
			${p_{r,k,{n_s}}}$                       & The allocated power of the link between RRH $r$
			\\[.2ex]
			$\text{p}_r^{\max }$    & Maximum transmit power of RRH $r$
			\\[.2ex]
			$\sigma_0^2$  & Noise variance
			\\[.2ex]
			${f_{{n_s},b}} \in \{0, 1\}$     & The association {factor} between the BBU $b$
			\\[.2ex]
			& and the user $n_s$
			\\[.2ex]
			 ${x_{{n_s},r}} \in \{0, 1\}$                 & The association {factor} between the RRH $r$
			 \\[.2ex]
			 & and the user $n_s$
			 \\[.2ex]
			 $\beta _{r,k,{n_s}} \in \{0, 1\}$   & User association variable
			 \\[.5ex]
			 \hline
					\end{tabular}
		\label{tabs} 
	\end{table}

	\subsection{Sensing Time Adjustment Problem}
		Energy detection method is used by RRHs for spectrum sensing due to its low computational and implementation
		complexities. Moreover, energy detector employs blind method to detect signals, i.e., there is no need to know primary users' signal and the output of the energy detector is compared with a threshold that depends on the noise floor. Regarding (\cite{liang2008sensing}, (67)), for a target probability of false alarm, ${\bar P_{fa}^k}$, detection probability at sub-carrier $k$ can be written as
		\begin{align}
			\label{detection probability}
			&{P_d^k} = \\ &\mathcal{Q}\left(\! {\frac{1}{\alpha_k }\left( {{\mathcal{Q}^{ - 1}}({{\bar P}_{fa}^k}) - \gamma_p \sum\limits_{r = 1}^R {\sqrt {{\tau _{r,k}}{\nu_{sa}}}{{\left| {{h_{r,k}^{HU}}} \right|}^2}} } \right)} \right) \text{~}\forall k \in \mathcal{K},\nonumber
		\end{align}
		where  ${h_{r,k}^{HU}}$ is the channel coefficient between RRH $r$ and the HVWN user at sub-carrier $k$ that is zero-mean, unit variance complex Gaussian random variable. Also, $\alpha_k  = \sqrt {2\gamma_p \sum\nolimits_{r = 1}^R {{{\left| {{h_{r,k}^{HU}}} \right|}^2} + 1} }$ and $\mathcal{Q}$, $\gamma_p$, and $\nu_{sa} $ are Q-function, received signal-to-noise ratio (SNR)
		of the HVWN user measured at the RRH, and sampling frequency, respectively.
	In order to preserve the HVWN user from the interference generated by LVWN users, the probability of detection $P_d^k$ should be {larger} than a target value $\bar{P_d}$. This constraint can be {expressed} as
	\begin{equation*}
		\label{cons2}
		\text{C1:~~}
		{P_d^k} \ge {{\bar P}_d,}  \text{~~~}\forall k \in \mathcal{K}.
	\end{equation*}
	 Let denote $M_{k} $ as the number of samples at the input of the energy detector. Therefore, for the given pair of target probabilities $(\bar{ P}_d,\bar P_{fa}^k) $, the minimum number of required samples to achieve these targets can be
	 obtained as follows \cite{liang2008sensing}
	 \begin{eqnarray}
	 \lefteqn{M_{k}^{\text{min}} =} \\&&\dfrac{4}{(\alpha^2_k - 1)^2 } \left[{\mathcal{Q}^{ - 1}}(\bar{P}_{fa}^k)- {\mathcal{Q}^{ - 1}}(\bar{ P}_d)\alpha_k\right]^2 \text{,~}\forall k \in \mathcal{K}. \nonumber
	 \end{eqnarray}
For each sensing node in the spectrum sensing, at least $M_{k}^{\text{min}}$ samples are required to insure target probabilities $(\bar{ P}_d,\bar P_{fa}^k) $. Indeed, if the sample size is not enough, the target probabilities will not hold. Clearly, in this circumstance referred to \textit{interruption}, LVWN users are not able to transmit while the spectrum is vacant.	
	
	  Furthermore, sensing time duration of RRH $r$ at sub-carrier $k$ should not exceed the frame time duration. This practical limitation can be represented as
	\begin{equation*}
		\label{cons1}
		\text{C2:~~}
		0 < {\tau _{r,k}} \le T,\ \text{~~~}\forall r \in \mathcal{R,}  \text{~~~}\forall k \in \mathcal{K}.
	\end{equation*}	
	\subsection{Resource Allocation Problem}
	As shown in Fig. 1, users of different slices are served by multiple RRHs connected to the BBU pool via front-haul links. Due to the cloud-based structure, a new set of variables, i.e., front-haul assignment parameters should be defined to mathematically present the structure of this setup. Moreover, the parameters and constraints of the resource allocation related to the network slicing and C-RAN are defined in this section.
	
	In this context, each user should be assigned to RRH and BBU. The association {factor} between the BBU $b$ and the user $n_s$ is
	represented by\\
	\begin{equation}
		\label{eq_5}
		{f_{{n_s},b}} = \left\{ {\begin{array}{*{20}{c}}
				\!\!\!1, \text{~~}\text{if user ${n_s}$ in slice $\mathcal{S}$ is supported by BBU $b$,}\\
				\!\!\!0, \text{~~}\text{otherwise,} \text{~~~~~~~~~~~~~~~~~~~~~~~~~~~~~~~~~~~~~~~~}
		\end{array}} \right.
	\end{equation}
and $\mathbf{f} ~=~ {\left[ {{f_{{n_s},b}}} \right]_{\forall b,s,{n_s}}}$ is the BBU association vector. Practically, each BBU $b$ can support at most $O^\text{max}$ users due to its processing limitations, i.e.,
	\begin{equation*}
		\label{Cons3}
		\text{C3:~~}
		\sum\limits_{s \in  \mathcal{S}} {\sum\limits_{{n_s} \in  \mathcal{N}_s} {{f_{{n_s},b}}} }  \le O^{\text{max} },  \text{~~~}\forall b \in \mathcal{B}.
	\end{equation*}
The association {factor} between the RRH $r$ and the user $n_s$ is
	represented {as}\\
	\begin{equation}
		\label{eq_6}
		{x_{{n_s},r}} = \left\{ {\begin{array}{*{20}{c}}
				\!\!\!1, \text{~~}\text{if user ${n_s}$ in slice $\mathcal{S}$ is connected to RRH $r$,}\\
				\!\!\!0, \text{~~}\text{otherwise.} \text{~~~~~~~~~~~~~~~~~~~~~~~~~~~~~~~~~~~~~~~~}
		\end{array}} \right.
	\end{equation}
	It is assumed that each user ${n_s}$ can only be connected to at most
	one RRH $r$ at each transmission frame, i.e.,
	\begin{equation*}
		\label{Cons4}
		\text{C4:~~}
		\sum\limits_{r \in \mathcal{R}}{x _{{n_s},r}} \le 1, \text{~~~~~~}\forall {n_s} \in \mathcal{N}_s,\text{~~~}\forall s \in \mathcal{S}.
	\end{equation*}
	To jointly consider user assignment to the RRHs and sub-carrier allocation, we define binary-valued $\beta _{r,k,{n_s}} \in \{0, 1\}$ as the user association variable (UAV). The value of UAV will be equal to one if RRH $r$ allocates sub-carrier $k$ to user $n_s$. OFDMA based sub-carrier allocation should be carried out for each cell exclusively, and it can be mathematically stated as
	\begin{equation*}
		\label{Cons5}
		\text{C5:~~} \sum\limits_{s \in \mathcal{S}} {\sum\limits_{{n_s} \in \mathcal{N}_s} {{\beta _{r,k,{n_s}}}} }  \le 1, \text{~~~}\forall r \in \mathcal{R,}  \text{~~~}\forall k \in \mathcal{K}.
	\end{equation*}
	Sub-carrier $k$ is allocated to user $n_s$ when that user is assigned to RRH $r$, this constraint can be represented by
	\begin{equation*}
		\label{Cons6}
		\text{C6:~~} {\beta _{r,k,{n_s}}} \le {x _{{n_s},r,}} \text{~~}\forall {n_s} \in \mathcal{N}_s,\text{~}\forall s \in \mathcal{S}\text{~}\forall r \in \mathcal{R,} \text{~}\forall k \in \mathcal{K}.
	\end{equation*}
	Since each front-haul link between RRH $r$ and BBU $b$ is limited and has a maximum capacity of $C_{r,b}^{\text{max} }$, 
	the following practical constraint should be considered in our setup
	\begin{equation*}
		\label{Cons7}
		\text{C7:~~}
		\sum\limits_{s \in \mathcal{S}}{\sum\limits_{{n_s} \in \mathcal{N}_s}  {{f_{{n_s},b}}{x_{{n_s},r}} } }  \le C_{r,b}^{\text{max} }, \text{~~~}\forall r \in \mathcal{R,}
		\text{~~~}\forall b \in \mathcal{B}.
	\end{equation*}
	Moreover, in order to avoid wasting of C-RAN resources and control the BBU load, each user is supported by only one BBU at each transmission instance, if and only if, it is assigned to at least one RRH. This constraint can be represented as
	\begin{equation*}
		\label{Cons8}
		\text{C8:~~} \sum\limits_{b \in \mathcal{B}}{f_{{n_s},b} = \sum\limits_{r \in \mathcal{R}}{x _{{n_s},r}} } , \text{~~}\forall {n_s} \in \mathcal{N}_s,\text{~~}\forall s \in \mathcal{S} .
	\end{equation*}
	Considering transmit power limitation of each RRH, we have
	\begin{equation*}
		\label{Cons9}
		\text{C9:~~} \sum\limits_{s \in \mathcal{S}} {\sum\limits_{{n_s} \in \mathcal{N}_s} {\sum\limits_{k \in \mathcal{K}} {{p_{r,k,{n_s}}}} } }  \le \text{p}_r^{\max }, \text{~~}\forall r \in \mathcal{R},
	\end{equation*}
	{where ${p_{r,k,{n_s}}}$ is the allocated power of the link between RRH $r$ and user $n_s$ at sub-carrier $k$, and  $\text{p}_r^{\max }$ is the maximum transmit power of RRH $r$.}
	
	It is assumed that the HVWN user is either active with the probability of $P^1$, or inactive with the probability of $P^0 = 1 - P^1$. Therefore, two scenarios for operating RRHs can be assumed. Indeed, RRHs can transmit when the HVWN user is either inactive and no false alarm is generated or active but not detected. Let $\gamma_{r,k,{n_s}}^0$ and $\gamma_{r,k,{n_s}}^1$ denote the SINR of the LVWN user $n_s$ when the corresponding RRH operates in the absence of HVWN user and when it operates in the presence of HVWN user, respectively.  Therefore, $\gamma_{b,k,{n_g}}^0$ and $\gamma_{b,k,{n_g}}^1$ can be expressed as
		
		\begin{equation}
			\label{Gamma0}
			\gamma_{r,k,{n_s}}^0 = \left( { \frac{{{p_{r,k,{n_s}}}{h_{r,k,{n_s}}}}}{{{\sigma _0^2} + I_{r,k,{n_s}}}}} \right),
		\end{equation}
		
		\begin{equation}
			\label{Gamma1}
			\gamma_{r,k,{n_s}}^1 = \left( { \frac{{{p_{r,k,{n_s}}}{h_{r,k,{n_s}}}}}{{{\sigma _0^2} + I_{r,k,{n_s}} + I_p}}} \right),
		\end{equation}
		where $h_{r,k,{n_s}}$ is the channel power gain of the link between RRH $r$ and user $n_s$ at sub-carrier $k$, and
	$
			I_{r,k,{n_s}} = \!\!\!\!\!\!\!\!\!\sum\limits_{\forall r' \in \mathcal{R},r' \ne r} {\sum\limits_{\forall s \in \mathcal{S}} {\sum\limits_{\forall {{n'}_s} \in \mathcal{N},{{n'}_s} \ne {n_s}} \!\!\!\!\!\!{{p_{r',k,{n'_s}}}{h_{r',k,{n_s}}}} } }
	$
		is induced interference to user $n_s$ in cell $r$ and sub-carrier $k$ from the other LVWN users. {Also, $\sigma _0^2$ and $I _p$ are the noise power and the interference of HVWN user measured at the LVWN receiver, respectively.} Resource allocation process for each RRH $r$ is performed every $T$ time frame consisting of the sensing {time} slot at each sub-carrier $k$, i.e., $\tau_{r,k}$, and the transmission {time} slot, i.e., $T-\tau_{r,k}$. Consequently, the average throughput {in bps/Hz} for a LVWN user can be stated as
	\begin{eqnarray}
		\label{eq10}
		\lefteqn{\Re_{r,k,{n_s}}(\boldsymbol{\beta},\boldsymbol{\tau} , \mathbf{p}) = \beta _{r,k,{n_s}} \frac{{T - \tau_{r,k} }}{T} \times}\\ && \Bigg[{P^0}{\log _2}\left( {1 + \gamma _{r,k,{n_s}}^0} \right)(1 - {P_{fa}^k})  \nonumber\\ && \,\,\,\,\,\,\,+   {P^1}{\log _2}\left( {1 + \gamma _{r,k,{n_s}}^1} \right)(1 - {P_d^k})\Bigg],\nonumber
	\end{eqnarray}
	where $\boldsymbol{\beta}$, $\boldsymbol{\tau}$ and \textbf{p} are, respectively, the vectors of all ${\beta _{r,k,{n_s}}}$, $\tau_{r,k}$, and ${p_{r,k,{n_s}}}$ for all $r \in \mathcal{R}$, $k \in \mathcal{K}$, ${n_s} \in \mathcal{N}_s$, and $ s \in \mathcal{S}$.
	In practical scenarios, desired detection probability is chosen to be close to but less than 1 \cite{liang2008sensing}. Also, to commercial justification for the secondary usage of the spectrum, we assume that the probability of being active for the HVWN users, i.e., $P^1$, is small. Based on the assumptions, the first term of (\ref{eq10}) is dominant, and for the different values of $P_{fa}^k$, the approximation of ${{\Re}_{r,k,{n_s}}}$, i.e., ${\tilde{\Re}_{r,k,{n_s}}}$ for all ${n_s} \in \mathcal{N}_s\text{~}\forall s \in \mathcal{S}\text{~}\forall r \in \mathcal{R} \text{~}\forall k \in \mathcal{K}$, can be expressed as
	\begin{eqnarray}
		\label{eq18}
		\lefteqn{\tilde{\Re}_{r,k,{n_s}}(\boldsymbol{\beta}, \boldsymbol{\tau} , \mathbf{p}) =}\\ &&\beta _{r,k,{n_s}} \frac{{T - \tau_{r,k} }}{T}\bigg[{P^0}{\log _2}\big( {1 + \gamma _{r,k,{n_s}}^0} \big) \nonumber
		\times(1 - {P_{fa}^k})\bigg]. 
	\end{eqnarray}
	To provide isolation between slices and support different throughput demands for the LVWN users, let us consider the minimum rate requirement of each slice $s \in \mathcal{S}$, $\Re_s^\text{rsv}$, which should be supported by RRHs, and therefore, we have the following constraint:
	\begin{equation*}
		\label{Cons11}
		\text{C10:~~~~}
		\sum\limits_{r \in \mathcal{R}}  {\sum\limits_{{n_s} \in \mathcal{N}_s} {\sum\limits_{k \in \mathcal{K}} {{\tilde{\Re}_{r,k,{n_s}}}(\boldsymbol{\beta}, \boldsymbol{\tau} , \mathbf{p})} } }   \ge \Re_s^\text{rsv}, \text{~~~}\forall s \in \mathcal{S}.
	\end{equation*}
	
	Now, the problem of  sensing time optimization, dynamic power and sub-carrier allocation as well as BBU and RRH assignment in a C-RAN empowered LVWN with the aim of throughput  maximization for the LVWN users can be formulated as
	\begin{equation}
		\label{eq12}
		\begin{aligned}
			& \underset{{\boldsymbol{\beta},\boldsymbol{\tau},\mathbf{p},\boldsymbol{f},\boldsymbol{x}}}{\text{max}}
			& & \sum\limits_{r \in \mathcal{R}} {\sum\limits_{s \in \mathcal{S}} {\sum\limits_{{n_s} \in \mathcal{N}_s} {\sum\limits_{k \in \mathcal{K}} {{\tilde{\Re}_{r,k,{n_s}}}(\boldsymbol{\beta}, \boldsymbol{\tau} , \mathbf{p})}}}} \\
			& \text{subject to}
			& & \text{C1 - C10}.
		\end{aligned}
	\end{equation}
	Due to {the} combination of continuous variables, e.g., $\mathbf{p}$ and $\boldsymbol{\tau}$ and the binary variables e.g., $\boldsymbol{\beta}$, $\mathbf{x}$ and  $\mathbf{f}$,  (\ref{eq12}) is a non-convex mixed-integer non-linear problem (MINLP), which is completely NP-hard and difficult to solve in practice \cite{burer2012non}. In the following sections, we design an efficient
	 algorithm to find a solution with reasonable computational complexity.
	\section{Proposed Algorithm}
	In order to tackle the computational complexity of (\ref{eq12}), a three-step iterative approach is proposed. Accordingly, sensing time, RRH as well as BBU association parameters, and power allocation are {optimized} in steps one, two, and three, respectively. More specifically, in each iteration $t$, step 1 computes the {optimized} value for sensing time $\boldsymbol{\tau}$ based on the given values of power allocation and association parameters from the previous iteration.
	In step 2, based on the given value of power allocation and the derived value of sensing time vector $\boldsymbol{\tau}(t)$ from step 1, the association problem is solved and the {optimized} values of $\mathbf{f}$, $\mathbf{x}$ and $\boldsymbol{\beta}$ are derived. {Then, for given} association parameters, i.e., $\mathbf{f}$, $\mathbf{x}$ and $\boldsymbol{\beta}$ and sensing time vector, i.e., $\boldsymbol{\tau}$, {power allocation problem is solved. }Iterative approaches for all 3 steps are executed {in alternation} until the current sensing time, association parameters and power allocation vector solutions {lie in slight deviation from the values} obtained in the previous iteration. 	In other words, the iterative procedure is stopped when the difference of optimal values of throughput in two consecutive iterations is less than $\varepsilon$ where $0 < \varepsilon  \ll 1$. The reduction of variables in each subproblem obviously provides both low computational complexity and more mathematical tractability for solving the original {problem}. The proposed analytical solutions for each subproblem are provided in details in the following subsections.
	\subsection{Step 1: Spectrum Sensing Problem }
	Given association parameters i.e., $\mathbf{f}(t-1)$, $\mathbf{x}(t-1)$ and $\boldsymbol{\beta}(t-1)$ and also $\mathbf{p}(t-1)$, at iteration $t$, \eqref{eq12} is transformed into
	\begin{equation}
		\label{sensing problem}
		\begin{aligned}
			& \underset{\boldsymbol{\tau}(t)}{\text{max}}
			& &
		\!\!\!\!	\sum\limits_{r \in \mathcal{R}} {\sum\limits_{s \in \mathcal{S}} {\sum\limits_{{n_s} \in \mathcal{N}_s} {\sum\limits_{k \in \mathcal{K}} {{\tilde{\Re}_{r,k,{n_s}}}(\boldsymbol{\beta}(t-1)}, {\tau(t)} ,\mathbf{p}(t-1))}}},\\
			& \text{s.t.:}
			& & \text{C1-C2,C10},
		\end{aligned}
	\end{equation}
	which only consists of $\boldsymbol{\tau}(t)$ as an optimum variable vector. While \eqref{sensing problem} is more simplified than \eqref{eq12}, still suffers from non-convexity. To convert \eqref{sensing problem} to a convex form, we define a new variable $\lambda_{r,k}=\sqrt{\tau_{r,k},\nu_{sa}}$ and rewrite \eqref{sensing problem} as
	\begin{equation}
		\label{sensing problem change tau}
		\begin{aligned}
			& \underset{\boldsymbol{\lambda}(t)}{\text{max}}
		\sum\limits_{r \in \mathcal{R}} {\sum\limits_{s \in \mathcal{S}} {\sum\limits_{{n_s} \in \mathcal{N}_s} {\sum\limits_{k \in \mathcal{K}} {{\tilde{\Re}_{r,k,{n_s}}}(\boldsymbol{\beta}(t-1)}, {\lambda(t)} ,\mathbf{p}(t-1))}}},\\
			&\text{s.t.:}\\
			& 			 \text{$\widehat{\text{C}}$1:~} \sum\limits_{r = 1}^R {{\lambda _{r,k}(t)}{{\left| {h_{r,k}^{PU}} \right|}^2}}  \geq \frac{1}{\gamma }\bigg[ {{Q^{ - 1}}\big( {{{P}_{fa}^k}} \big) - {\alpha _k}{Q^{ - 1}}\big( {{{\bar P}_d}} \big)} \bigg],\\
						&  \text{~~~~~}\forall k \in \mathcal{K},\\
			&
			 \text{$\widehat{\text{C}}$2:~} 0 < {\lambda _{r,k}(t)} \le \sqrt{T\nu_{sa}},\ \text{~~~}\forall r \in \mathcal{R},  \text{~~~}\forall k \in \mathcal{K},\\
						&
			 \text{$\widehat{\text{C}}$10:~}
			\sum\limits_{r \in \mathcal{R}}  {\sum\limits_{{n_s} \in \mathcal{N}_s} {\sum\limits_{k \in \mathcal{K}} {{\tilde{\Re}_{r,k,{n_s}}}(\boldsymbol{\beta}(t-1)}, {\lambda(t)} ,\mathbf{p}(t-1))}}   \ge \Re_s^\text{rsv},\\
			&\text{~~~~~~~}\forall s \in \mathcal{S}. \\
		\end{aligned}
	\end{equation}
{As the objective function and left side of $\widehat{\text{C}}10$ are concave functions in terms of ${\lambda_{r,k}}$, and the constraints $\widehat{\text{C}}1$ and $\widehat{\text{C}}2$ are linear, problem \eqref{sensing problem change tau} is a convex optimization problem and can be solved optimally using either lagrangian method or software packages, e.g., CVX \cite{cvx}.}
	\subsection{Step 2: User Association Problem}
	For a fixed value of $\boldsymbol{\tau}$ obtained from Step 1 and given $\mathbf{p}(t-1)$, at iteration $t$, \eqref{eq12} is transformed into an integer non-linear problem (INLP) as follows\\
	\begin{equation}
		\label{Association-Problem}
		\begin{aligned}
			& \underset{{\boldsymbol{\beta}(t),\boldsymbol{f}(t),\boldsymbol{x}(t)}}{\text{max}}
			& & \!\!\!\!\sum\limits_{r \in \mathcal{R}} {\sum\limits_{s \in \mathcal{S}} {\sum\limits_{{n_s} \in \mathcal{N}_s} {\sum\limits_{k \in \mathcal{K}} {{\tilde{\Re}_{r,k,{n_s}}}({\beta(t),\boldsymbol{\tau}(t),\mathbf{p}(t-1)})}}}} \\
			& \text{~~~~~~s.t.:}
			& & \text{C3-C8,C10}.
		\end{aligned}
	\end{equation}
By converting constraint C7 into a linear form, problem \eqref{Association-Problem} will be transformed into the integer linear programming (ILP), and can be solved via slover software packages e.g., CVX with the internal solver MOSEK \cite{mosek}. Let introduce $y_{b,r,{n_s}} = f_{{n_s},b}x_{{n_s},r}$ and $y_{b,r,{n_s}} \in \{0,1\}$. Therefore, C7 is transformed into
	\begin{equation*}
		\begin{aligned}
			&\text{$\widehat{\text{C}}$7.1}:
			& & \sum\limits_{s \in \mathcal{S}}{\sum\limits_{{n_s} \in \mathcal{N}_s}  {{y_{b,r,{n_s}}} } }  \le C_{r,b}^{\text{max} }, \text{~~}\forall r \in \mathcal{R,}
			\text{~~~}\forall b \in \mathcal{B,}\\
			&\text{$\widehat{\text{C}}$7.2}:
			& & {{y_{b,r,{n_s}}} }  \le {f_{{n_s},b}} , \text{~}\forall {n_s} \in \mathcal{N}_s,\text{~}\forall s \in \mathcal{S,}\text{~}\forall r \in \mathcal{R,} \text{~}\text{~}\forall b \in \mathcal{B,}\\
			&\text{$\widehat{\text{C}}$7.3}:
			& & {{y_{b,r,{n_s}}} }  \le {x_{{n_s},r}} , \text{~}\forall {n_s} \in \mathcal{N}_s,\text{~}\forall s \in \mathcal{S,}\text{~}\forall r \in \mathcal{R,} \text{~}\text{~}\forall b \in \mathcal{B,}\\
			&\text{$\widehat{\text{C}}$7.4}:
			& & {{y_{b,r,{n_s}}} } \ge  {f_{{n_s},b}} + {x_{{n_s},r}} -1 ,\\
				&
			& & \text{~}\forall {n_s} \in \mathcal{N}_s,\text{~}\forall s \in \mathcal{S,}\text{~}\forall r \in \mathcal{R,} \text{~}\text{~}\forall b \in \mathcal{B}.\\
		\end{aligned}
	\end{equation*}
	As a result, \eqref{Association-Problem} can be rewritten as follows
	\begin{equation}
		\label{Association-Problem-convert}
		\begin{aligned}
			& \underset{{\boldsymbol{\beta}(t),\boldsymbol{f}(t),\boldsymbol{x}(t),\boldsymbol{y}(t)}}{\text{max}}
		\!\!\!\!\!\!\!	& & \sum\limits_{r \in \mathcal{R}} {\sum\limits_{s \in \mathcal{S}} {\sum\limits_{{n_s} \in \mathcal{N}_s} {\sum\limits_{k \in \mathcal{K}} {{\tilde{\Re}_{r,k,{n_s}}}({\beta(t),\boldsymbol{\tau}(t),\mathbf{p}(t-1)})}}}} \\
			& \text{~~~~~~~s.t.:}
			& & \text{C3-C6},\text{$\widehat{\text{C}}$7.1-}\text{$\widehat{\text{C}}$7.4},\text{C8,C10}
		\end{aligned}
	\end{equation}
	which belongs to ILP.
	\subsection{Step 3: Power Allocation Problem}
	For the fixed values of $\boldsymbol{\tau}$ and $\boldsymbol{\beta}$ obtained from Step 1 and 2, at iteration $t$, power allocation problem can be written as \\
	\begin{equation}
		\label{Power alloction problem}
		\begin{aligned}
			& \underset{\mathbf{p}(t)}{\text{max}}
		\!\!\!\!\!\!\!\!\!	& &\sum\limits_{r \in \mathcal{R}} {\sum\limits_{s \in \mathcal{S}} {\sum\limits_{{n_s} \in \mathcal{N}_s} {\sum\limits_{k \in \mathcal{K}} {{\tilde{\Re}_{r,k,{n_s}}}(\boldsymbol{\beta}(t),\boldsymbol{\tau}(t),{p}(t))}}}}\\
			& \text{~s.t.:}
			& & \text{C9-C10.}
		\end{aligned}
	\end{equation}
	Since the objective function and C10 are non-concave functions, \eqref{Power alloction problem} is a non-convex optimization problem. Our approach for solving \eqref{Power alloction problem} is to convert it into a convex optimization problem by employing the {successive convex approximation} (SCA) algorithm where  ${\tilde{\Re}_{r,k,{n_s}}}(\mathbf{p})$ is converted into a concave form based
	on the difference-of-two-concave-functions (D.C.) approximation method \cite{tuan2000dc}. Accordingly, at each iteration $l$ of the SCA algorithm, we define ${\tilde{\Re}_{r,k,{n_s}}}(\mathbf{p}(l)) = {u_{r,k,{n_s}}}(\mathbf{p}(l)) - {v_{r,k,{n_s}}}(\mathbf{p}(l))$ where  ${u_{r,k,{n_s}}}(\mathbf{p}(l))$ and ${v_{r,k,{n_s}}}(\mathbf{p}(l))$ are the concave functions as
	\begin{align}
		{u_{r,k,{n_s}}}(\mathbf{p}(l)) =& \beta _{r,k,{n_s}}\frac{{T - \tau_{r,k} }}{T}{P^0}(1 - {P_{fa}^k})  \\
		&\times {\log _2}\big[ {{\sigma _0^2} + I_{r,k,{n_s}} + {p_{r,k,{n_s}}}(l){h_{r,k,{n_s}}}}\big],\nonumber
	\end{align}
	\begin{align}
		{v_{r,k,{n_s}}}(\mathbf{p}(l)) = {\beta _{r,k,{n_s}} \frac{{T - \tau_{r,k} }}{T}{P^0}(1 - {P_{fa}^k}){\log _2}\left[ {{\sigma _0^2} + I_{r,k,{n_s}}}\right],}
	\end{align}
	where $\mathbf{p}(l)$ is the power allocation vector at iteration $l$. Also, at each iteration $l$, ${v_{r,k,{n_s}}}(\mathbf{p}(l))$ is replaced with its first-order Taylor expansion as
	\begin{align}
		\label{D.C. Approximation}
		{v_{r,k,{n_s}}}(\mathbf{p}(l)) \approx& {v_{r,k,{n_s}}}(\mathbf{p}(l-1)) \\ &+ \nabla{{v_{r,k,{n_s}}}(\mathbf{p}(l-1))}(\mathbf{p}(l)-\mathbf{p}(l-1)),\nonumber
	\end{align}
where	$\nabla{{v_{r,k,{n_s}}}(\mathbf{p}(l-1))}$  is the gradient vector of ${v_{r,k,{n_s}}}(\mathbf{p}(l-1))$ with respect to the vector $\mathbf{p}(l-1)$ and its element, can be obtained by
	\begin{equation}\label{gg nabla}
		\nabla{{v_{r,k,{n_s}}}(\mathbf{p}(l-1))}  =
		\left\{
		\begin{array}{ll}
		\!\!\!	0, & \hbox{$ r=r', n_s=n'{_s}$}, \\
			\!\!\!\frac{h_{r,k,n_s}  }{ \ln2 (I_{r,k,n_s} +  {\sigma _0^2})  }, & \hbox{$\forall r \neq r', n_s\neq n'{_s}$}
		\end{array}
		\right.
	\end{equation}
where $ r' \ne r \in \mathcal{R}$.	Therefore, at each iteration $l$, approximated concave throughput, i.e., ${\hat \Re_{r,k,{n_s}}}$, can be stated as
	\begin{align}
		\label{conv-approx}
		{\hat \Re_{r,k,{n_s}}(\mathbf{p}(l))} \approx& {u_{r,k,{n_s}}}(\mathbf{p}(l)) - {v_{r,k,{n_s}}}(\mathbf{p}(l-1))  \\ &-\nabla{{v_{r,k,{n_s}}}(\mathbf{p}(l-1))}(\mathbf{p}(l)-\mathbf{p}(l-1)).\nonumber
	\end{align}
	Hence, by substituting ${\hat \Re_{r,k,{n_s}}}$ in (\ref{Power alloction problem}), the convex
	approximated problem at each iteration $l$ is derived as follows:
	\begin{equation}
		\label{Approximation of Power alloction problem}
		\begin{aligned}
			& \underset{\mathbf{p}(l)}{\text{max}}
			& &\sum\limits_{r \in \mathcal{R}} {\sum\limits_{s \in \mathcal{S}} {\sum\limits_{{n_s} \in \mathcal{N}_s} {\sum\limits_{k \in \mathcal{K}} {\hat \Re_{r,k,{n_s}}(\mathbf{p}(l))}}}}\\
			& \text{s.t.:}
			& & \text{C9:~}
			\sum\limits_{s \in \mathcal{S}} {\sum\limits_{{n_s} \in \mathcal{N}_s} {\sum\limits_{k \in \mathcal{K}} {{p_{r,k,{n_s}}}}(l) } }  \le \text{p}_r^{\max }, \text{~}\forall r \in \mathcal{R} \\
			&
			& & \text{$\widehat{\text{C}}$10:~}
			\sum\limits_{r \in \mathcal{R}}  {\sum\limits_{{n_s} \in \mathcal{N}_s} {\sum\limits_{k \in \mathcal{K}} {\hat \Re_{r,k,{n_s}}(\mathbf{p}(l))} } }   \ge \Re_s^\text{rsv}, \text{~}\forall s \in \mathcal{S}. \\
		\end{aligned}
	\end{equation}
		\\\textbf{Preposition 1}: By employing D.C. approximation (\ref{D.C. Approximation}), (\ref{Approximation of Power alloction problem}) will finally converge to a locally optimal solution $\mathbf{p^*}$ of (\ref{Power alloction problem}).
	\\\textit{Proof}. Please see Appendix.

{Algorithm 1 represents an algorithm that solves (\ref{eq12}) to find $\tau (t)$, $f(t)$, $\beta (t)$, $x(t)$ and $p(t)$ in alternation.}
	\begin{algorithm}
		\label{algorithm-DC}
		\textbf{Initialization}:	Set $0 < \varepsilon  \ll 1$ and also  initial points $f(0)$, $\beta (0)$, $x(0)$, $p(0)$ and $t=1$;  \\
		\textbf{Repeat}:	\\
		\textbf{Step 1:} Given $f(t-1)$, $\beta (t-1)$, $x(t-1)$, $p(t-1)$, solve (\ref{sensing problem}) to find $\tau (t)$;\\
		\textbf{Step 2:} Given $\tau (t)$ and $p(t-1)$, solve (\ref{Association-Problem}) to find $f(t)$, $\beta (t)$ and $x(t)$ ;\\
        \textbf{Step 3:} Given $\tau (t)$, $f(t)$, $\beta (t)$ and $x(t)$, solve (\ref{Power alloction problem}) to find $p(t)$ ;\\
		\textbf{Step 4:} If $\left\| {\tilde{\Re}_{r,k,{n_s}}}(t) - {\tilde{\Re}_{r,k,{n_s}}}(t-1) \right\| \le \varepsilon$, stop. Otherwise, $t=t+1$ and go back to Step 1.
			\caption{Alternating algorithm for problem (\ref{eq12})}
\end{algorithm}
Also, for solving  (\ref{Power alloction problem}) in Step 3 of Algorithm 1, the general DC-based iterative algorithm is given in Algorithm 2 in which the disciplined convex programming (DCP) problem (\ref{Approximation of Power alloction problem}) can be solved by using the available optimization toolboxes, such as CVX.	
		\begin{algorithm}
			\label{algorithm-DC}
			\caption{DC-Based Algorithm For Solving Power Allocation Problem}
			\textbf{Initialization}:	Set $0 < \zeta  \ll 1, l=0$ and for a given $\mathbf{p}(l=0)$ as any feasible power vector;\\
			\textbf{Iterations}:	For a given $\mathbf{p}(l)$, execute three steps below;\\
			\textbf{Step 1:} Compute convex approximation from (\ref{conv-approx});\\
			\textbf{Step 2:} Solve (\ref{Approximation of Power alloction problem}) to obtain an optimal solution of $\mathbf{p}(l)$ ;\\
			\textbf{Step 3:} If $\left\| \mathbf{p}(l) - \mathbf{p}(l-1) \right\| \le \zeta$, stop. Otherwise, $l=l+1$ and go back to Step 1.
		\end{algorithm}
	\section{Computational Complexity and Convergence}
	In this section, two important aspects to evaluate the performance of the proposed algorithm, i.e., the computational complexity and convergence analysis are provided.
	\subsection{Computational Complexity}
	To analyze total computational complexity of the proposed algorithm, the complexity order of each step should be investigated. Step 1 consists of the calculation of the detection probability and the comparison of achieved throughput with the reserved rate for each slice, which has the complexity of $O(K+S)$. Moreover, in the second and third Step{s}, CVX is used to solve the binary user association problem (\ref{Association-Problem-convert}) and the power allocation problem (\ref{Approximation of Power alloction problem}) via interior point method (IPM). The computational complexity for this method is $\frac{{\log \left( {{\raise0.7ex\hbox{$c$} \!\mathord{\left/
						{\vphantom {c {{t^0}\alpha }}}\right.\kern-\nulldelimiterspace}
					\!\lower0.7ex\hbox{${({i^0}\alpha )}$}}} \right)}}{{\log \left( \mu  \right)}}$, where
	$c$ is the total number of constraints, $i^0$ is the initial point for
	approximating the accuracy of the IPM, $0 < \alpha  \ll 1$
	is the stopping criterion for the IPM, and $\mu$ is used
	for updating the accuracy of the IPM \cite{boyd2004convex}. The number of constraints in (\ref{Association-Problem-convert}) are $(N_sS(2+RK+3RB)+R(B+K)+B+S)$, where (\ref{Approximation of Power alloction problem}) has $(R+S)$ constraints. Clearly, one can realize that the complexity of the proposed scheme has polynomial growth which is a major improvement
	over direct search methods with exponential complexities.
	\subsection{Convergence}
	Our proposed algorithm for solving (\ref{eq12}) is based on block coordinate descent method (BCD) \cite{razaviyayn2013unified}. More specifically in this method, at each iteration, the objective function of (\ref{eq12}) is maximized in respect of a single block of variables while the other variables are held fixed. The convergence of the BCD method is guaranteed when the solution of each subproblem at each iteration brings global optimum. Moreover, in \cite{razaviyayn2013unified}, a unified convergence analysis for a general class of inexact BCD methods is provided when a sequence of approximate versions of the original problem is solved successively, and the convergence condition of this alternative inexact BCD approach is provided. In our proposed algorithm, both sensing and user association problems attain optimal solution while D.C. approximation of power allocation problem converges to a locally optimal solution. Thus, the convergence of the proposed algorithm to a local optimal solution is guaranteed, however, it may not be the global optimum.
\section{Simulation Results}
	To evaluate sensing-throughput tradeoff in our setup as well as verify our proposed algorithm, simulation results are provided in this section. We consider a multi-cell VWN with $R = 4$ RRHs connected to $B = 3$ BBUs which are serving $N_s = 8$ users in $S = 2$ slices in a 2km$\times$2km square area. We also assume that the total bandwidth is divided into $K = 16$ sub-carriers.  Using uniform distribution, the users are {placed} randomly inside of the square area and the coordinates of RRHs are: $(0.5$km, $0.5$km), $(0.5$km, 1.5km), (1.5km, 0.5km), (1.5km, 1.5km). The channel power gain is modeled by assuming large and small scale
	fading as $h_{r,k,n_s} = \psi_{r,k,n_s}d_{r,n_s}^{-a}$, where $a = 3$ is the path loss exponent, $d_{r,n_s}>0$ is the distance between RRH $r$ and user $n_s$ and $\psi_{r,k,n_s}$ is the exponential random variable with mean of 0.5 \cite{goldsmith2005wireless}. For the sake of simplicity, when the RRHs are used to sense the vacant channel, we assume that the received SNRs from HVWN user at each RRH are all equal, i.e., $\gamma_p = -15$dB \cite{liang2008sensing}. For all of the simulations $\zeta$ and $\epsilon$ are set to $10^{-3}$. In addition, we set $\Re_s^\text{rsv} = \Re^\text{rsv} = 4$ $\forall s \in \mathcal{S}$, ${\bar{P}_{fa}^k}=P_{fa~} \forall k \in \mathcal{K}$, and  maximum transmission power $\text{p}_r^{\max } = \text{p}^{\max }= 30$dBm  $\forall r \in \mathcal{R} $.

As we mentioned above, the achieved throughput of the LVWN is a function of sensing time $\tau$. Therefore, sensing time must be choose optimally to increase the vacant spectrum utilization by increasing the duration of transmission time slot as much as possible and also avoiding interruption. To investigate this issue, in
Fig. \ref{fig_2_outage}, the interruption probability versus $\tau$ for different values of $\bar{P_d}$ and ${P_{fa}} = 0.2$ is presented. We notice that the interruption probability decrease via increasing sensing time, and for a fixed sensing time, because of the lower number required for samples, the lower target detection probability yields the less interruption probability.
		\begin{figure}[!t]
		\centering
		\includegraphics[width=3.4in]{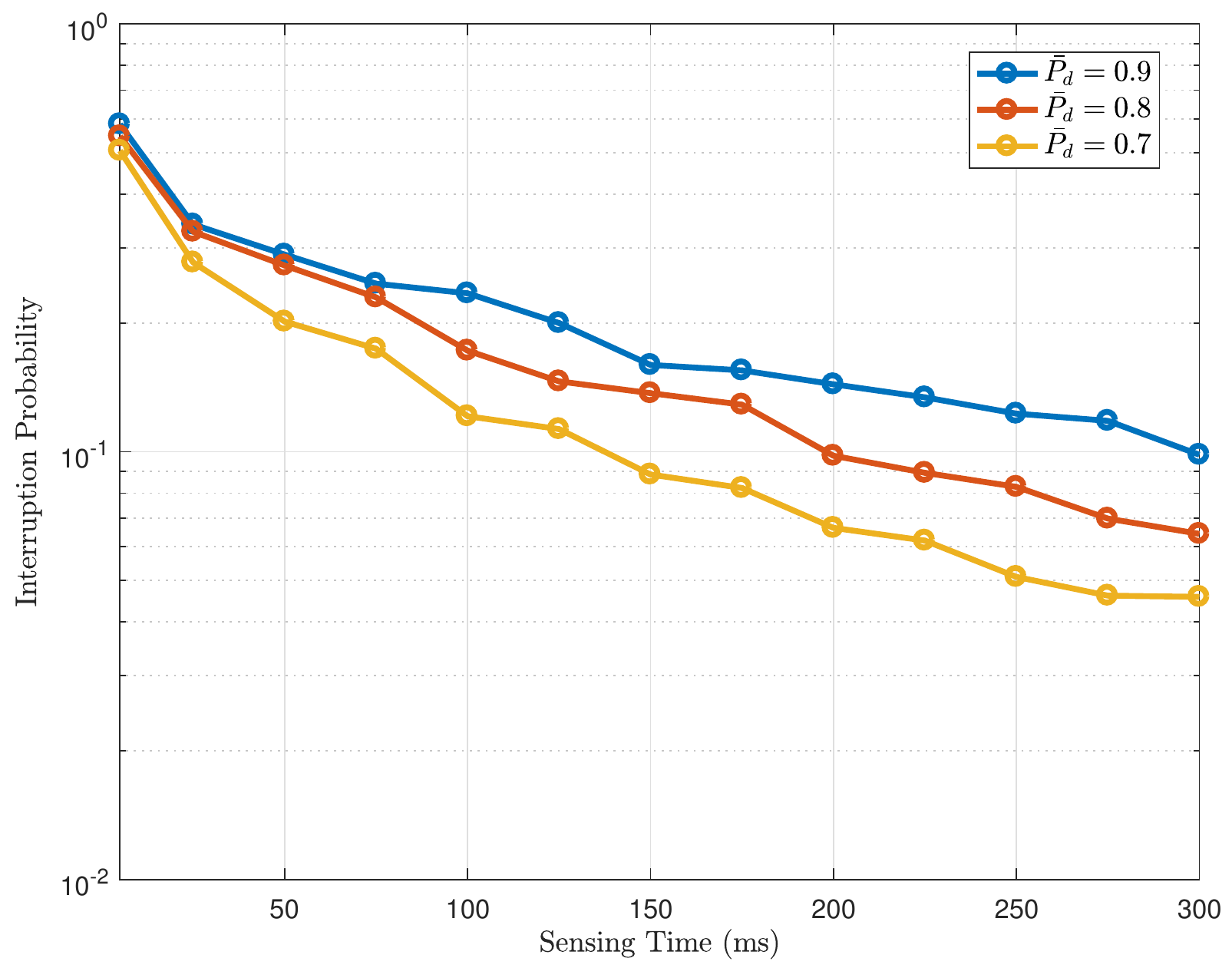}
		\caption{Interruption  Probability for LVWN users versus sensing time and $\bar{P_d}$ when $P_{fa} = 0.2$}
		\label{fig_2_outage}
	\end{figure}

To obtain more insight into the importance of optimizing $\tau$, the achievable throughput versus the fixed sensing time allocated to each transmission frame for the LVWN users is demonstrated in Fig. \ref{fig_2}. Obviously, for a given pair of target probabilities ($\bar{P_d}, P_{fa}$), one can conclude that there exists an optimal sensing time in which the value of the achievable throughput for the LVWN is maximized. Also, it can be seen that the optimum value of sensing time decreases with decreasing target detection probability $\bar{P_d}$. For instance, for $\bar{P_d} = 0.9 \text{~and~} 0.8$, the maximum value of achievable throughput is achieved at $\tau = 159$ ms and $\tau = 148$ ms, respectively. It is worth mentioning that since the QoS of the HVWN user will be guaranteed by choosing the higher probability of detection, decreasing $\bar{P_d}$ results in more throughput for the LVWN at the expense of more interference with the HVWN.  Figures \ref{fig_2_outage} and \ref{fig_2} highlight that choosing optimal sensing time significantly improves network performance.
		\begin{figure}[!t]
		\centering
		\includegraphics[width=3.3in]{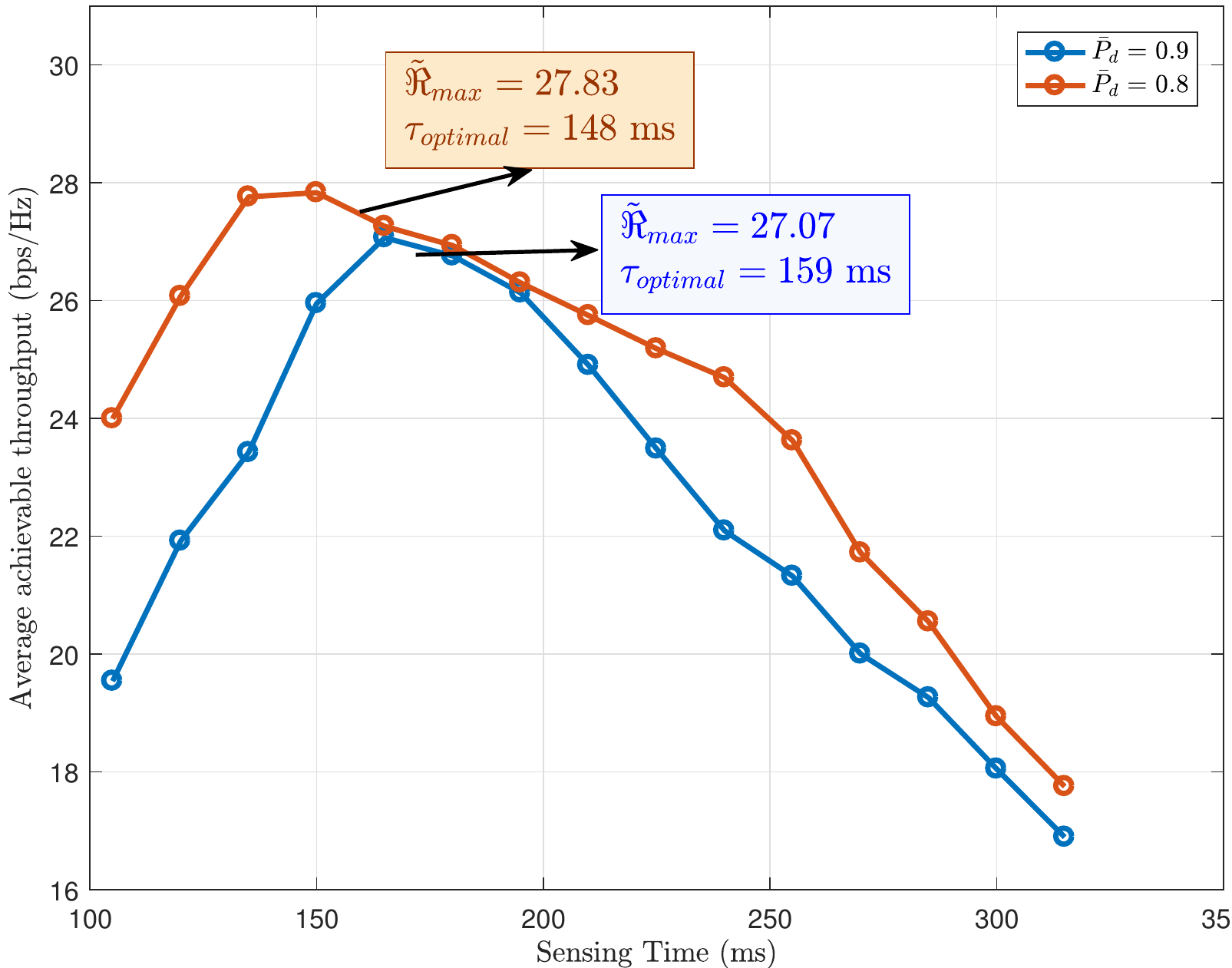}
		\caption{Average achievable throughput for LVWN versus sensing time for different $\bar{P_d}$ when $P_{fa} = 0.2$}
		\label{fig_2}
	\end{figure}

To study the effect of $\bar{P_d}$ and $P_{fa}$ as well as other system parameters on the optimal value of $\tau$, first, in Fig. \ref{fig2}, the optimal values of sensing time versus $\bar{P_d}$ and $P_{fa}$ is presented. From Fig. \ref{fig2}, we can conclude that at a certain value of detection probability, the optimal value of sensing time decrease with increasing $P_{fa}$ which leads to the shorter transmission time slot, i.e., less throughput for LVWN users. However, the lower the false alarm, the higher probability of successful spectrum utilization for the LVWN users. This can also be realized from (\ref{eq18}) which states that the average of achievable throughput for the LVWN is directly related to $(1 - P_{fa})$ and decreasing $P_{fa}$ yields more throughput for the network.
		\begin{figure}[!t]
		\centering
		\includegraphics[width=3.4in]{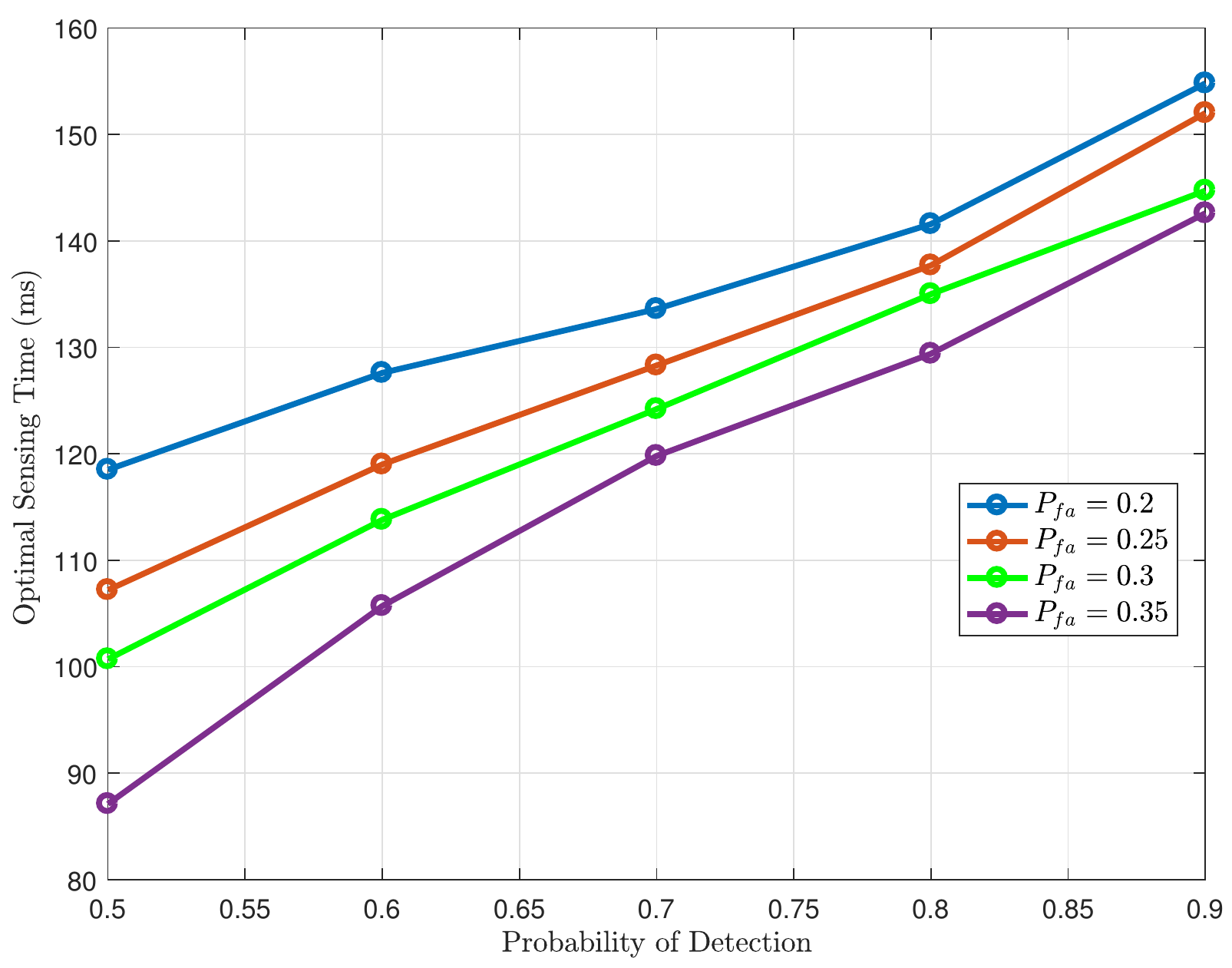}
		\caption{Optimal sensing time versus detection probability for the different values of false alarm probability}
		\label{fig2}
	\end{figure}

Fig. \ref{fig4} plots the average of achievable throughput of the LVWN versus number of users for optimal sensing time and the fixed sensing time scenario where the sensing time has been chosen large enough to ensure the target probabilities for this case. As we expected from user diversity gain, with increasing the number of users in LVWN, its total throughput increases. Our proposed approach to dynamically optimize sensing time outperforms the fixed sensing time scenario considerably. In a nutshell, Figs. \ref{fig2} and \ref{fig4} demonstrate sensing-throughput tradeoff in this setup.
			\begin{figure}[!t]
			\centering
			\includegraphics[width=3.4in]{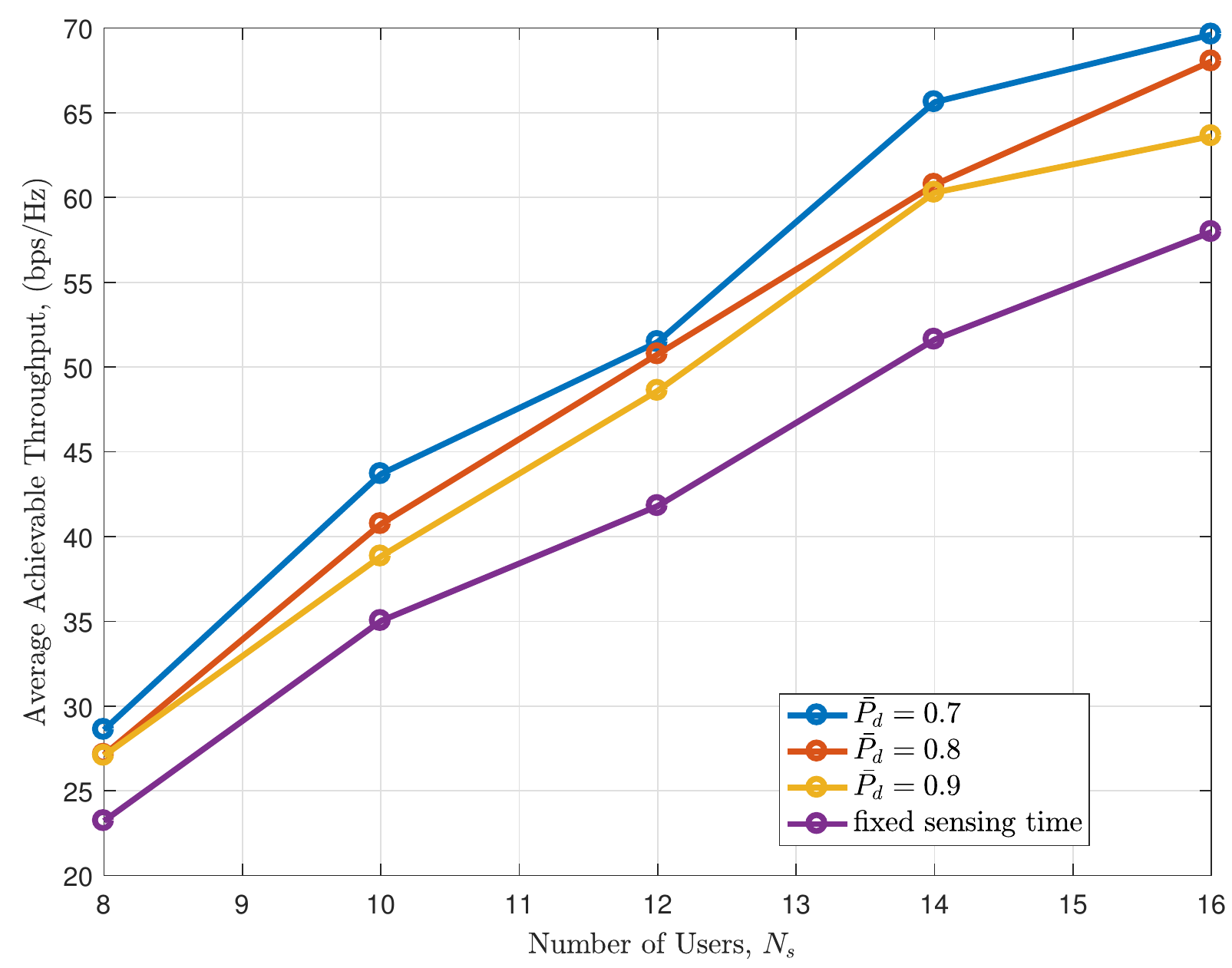}
			\caption{Average achievable throughput versus the number of users and the different values of $\bar{P_d}$ when $P_{fa} = 0.2$}
			\label{fig4}
		\end{figure}

	To highlight the importance of deploying C-RAN architecture, we investigate the effect of utilizing cooperative spectrum sensing,  on the optimum value of sensing time. From Fig. \ref{fig5} we can observe that, at a specified value of $\bar{P_d}$, the optimal value of sensing time decreases when the number of RRHs increases. Clearly, increasing the number of RRHs helps to mitigate the hidden node problem  at the {expense} of increasing processing complexity in terms of massage passing between nodes. However, in contrast to the traditional RAN, by using centralized powerful BBUs in C-RAN, this type of massage passing is significantly decreased.
\begin{figure}[!t]
	\centering
	\includegraphics[width=3.4in]{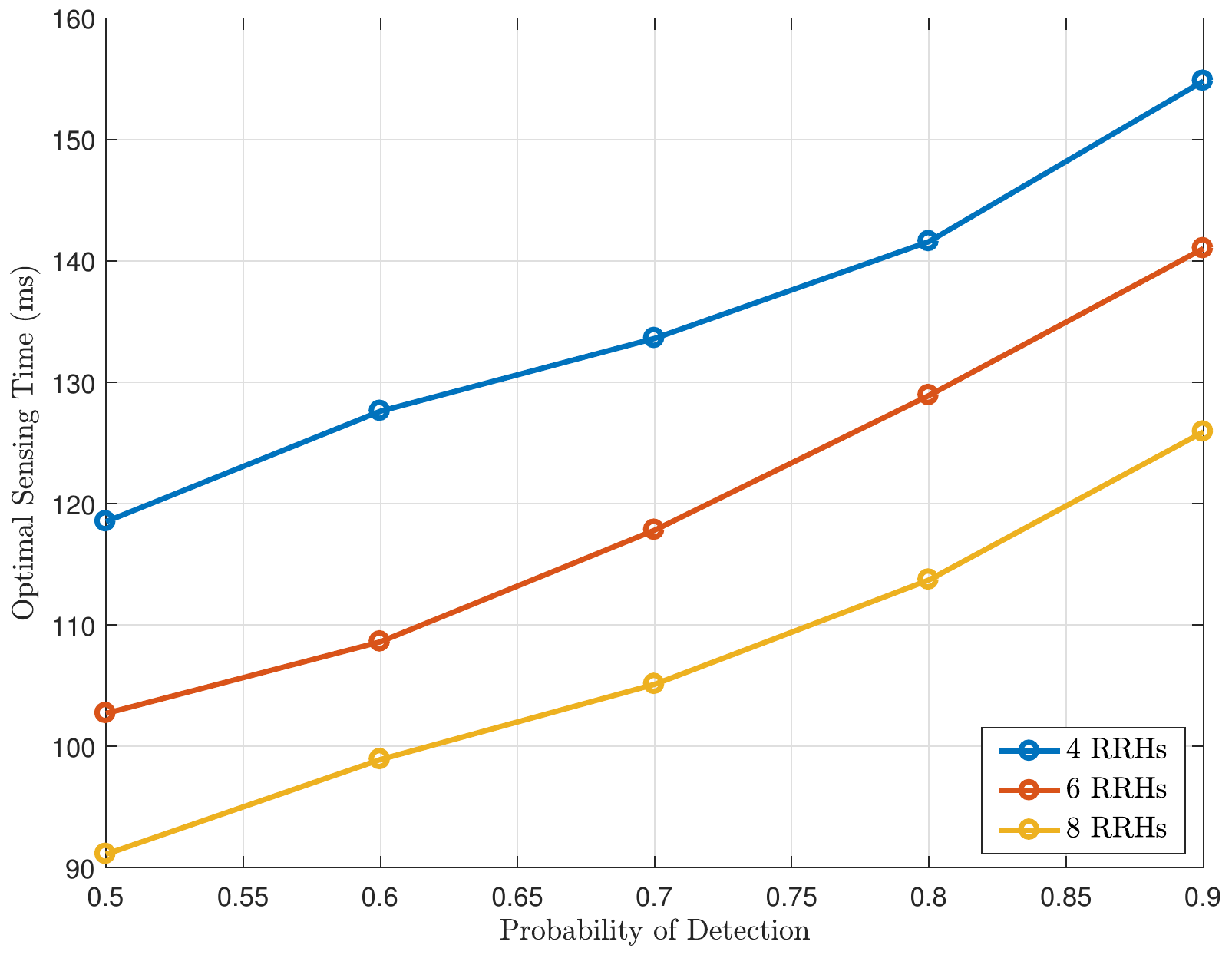}
	\caption{ Optimal sensing time (ms) for different number of RRHs, ${P_{fa}} = 0.2$ }
	\label{fig5}
\end{figure}
	\section{Concluding Remarks and Future Direction}
	In this paper, by utilizing the cooperative spectrum sensing empowered by C-RAN, we introduce a new approach for opportunistic spectrum sharing. Since the interests of HVWN users and LVWN users are contradictory, to avoid interference as well as optimal use of C-RAN resources, we propose the concept of sensing-throughput tradeoff and present a cloud-based spectrum sensing and dynamic resource allocation in a C-RAN empowered slice-based 5G network. We then formulate the joint sensing time, power and sub-carrier allocation and RRH assignment for LVWN{s} in fronthaul-limited C-RAN with the aim of throughput maximization. To solve this non-convex and NP-hard optimization problem, we deploy a three-step iterative algorithm where the first and second steps are optimally solved and the latter is solved by employing SCA approach. {Moreover, the complexity analysis and convergence behavior of our proposed algorithm are investigated and the algorithm has been evaluated by the simulation results.}
	The future directions of this work involve investigating heterogeneous nodes in terms of their different sensing capabilities and simultaneous multi-band spectrum sensing.
	\appendix
	\section{Proof of Preposition 1}
	${v_{r,k,{n_s}}}(\mathbf{p})$ is a concave function, hence, at each iteration $l$ we have
	\begin{align}
		\label{D.C. Approximation Appendix}
		{v_{r,k,{n_s}}}(\mathbf{p}(l)) \le &{v_{r,k,{n_s}}}(\mathbf{p}(l-1)) \nonumber\\ &\!\!\!+ \nabla{{v_{r,k,{n_s}}}(\mathbf{p}(l-1))}(\mathbf{p}(l)-\mathbf{p}(l-1)).
	\end{align}
	Moreover, by using (\ref{conv-approx}), ${\tilde \Re_{r,k,{n_s}}(\mathbf{p}(l))}$ is approximated to a concave form. Hence, it follows that
	\begin{multline}
			\label{rate-slice-appendix}
	\!\!\!\!\!\!	\sum\limits_{r \in \mathcal{R}}  \sum\limits_{{n_s} \in \mathcal{N}_s} {\sum\limits_{k \in \mathcal{K}}} \bigg[{{u_{r,k,{n_s}}}(\mathbf{p}(l))-{v_{r,k,{n_s}}}(\mathbf{p}(l-1))-} \\
	\begin{aligned}
	& { \!\!\!\nabla{{v_{r,k,{n_s}}}(\mathbf{p}(l-1))}(\mathbf{p}(l)-\mathbf{p}(l-1))} \bigg]     \ge \Re_s^\text{rsv}, \text{~}\forall s \in \mathcal{S}.\\
	\end{aligned}
	\end{multline}
	According to (\ref{D.C. Approximation Appendix}) and (\ref{rate-slice-appendix}), we have
	\begin{equation}
		\label{rate-slice-appendix-next-itr}
		\sum\limits_{r \in \mathcal{R}}  {\sum\limits_{{n_s} \in \mathcal{N}_s} {\sum\limits_{k \in \mathcal{K}} \{{{u_{r,k,{n_s}}}(\mathbf{p}(l))-{v_{r,k,{n_s}}}(\mathbf{p}(l)) } } }\}   \ge \Re_s^\text{rsv}, \text{~}\forall s \in \mathcal{S}.
	\end{equation}
	Accordingly, optimal solution at each iteration is always feasible. Furthermore, it can be easily shown that
	\begin{align*}
		\sum\limits_{r \in \mathcal{R}} {\sum\limits_{s \in \mathcal{S}} {\sum\limits_{{n_s} \in \mathcal{N}_s} {\sum\limits_{k \in \mathcal{K}} {{u_{r,k,{n_s}}}(\mathbf{p^*}(l))-{v_{r,k,{n_s}}}(\mathbf{p^*}(l))}}}}
	\end{align*}
	\begin{multline*}
		\ge 		\sum\limits_{r \in \mathcal{R}} {\sum\limits_{s \in \mathcal{S}}} {\sum\limits_{{n_s} \in \mathcal{N}_s}} {\sum\limits_{k \in \mathcal{K}}} \bigg[{{u_{r,k,{n_s}}}(\mathbf{p^*}(l))-{v_{r,k,{n_s}}}(\mathbf{p}(l-1))}\\
	\begin{aligned}
	& -\nabla{{v_{r,k,{n_s}}}(\mathbf{p}(l-1))}(\mathbf{p^*}(l)-\mathbf{p}(l-1))\bigg]\\
	\end{aligned}
	\end{multline*}
	\begin{multline*}
		= \underset{\mathbf{p}(l)}{\text{max}}		\sum\limits_{r \in \mathcal{R}} {\sum\limits_{s \in \mathcal{S}}} {\sum\limits_{{n_s} \in \mathcal{N}_s}} {\sum\limits_{k \in \mathcal{K}}} \bigg[{{u_{r,k,{n_s}}}(\mathbf{p}(l))-{v_{r,k,{n_s}}}(\mathbf{p}(l-1))} \\
\begin{aligned}
& -\nabla{{v_{r,k,{n_s}}}(\mathbf{p}(l-1))}(\mathbf{p}(l)-\mathbf{p}(l-1))\bigg]\\
\end{aligned}
\end{multline*}
	\begin{multline*}
\ge 		\sum\limits_{r \in \mathcal{R}} {\sum\limits_{s \in \mathcal{S}}} {\sum\limits_{{n_s} \in \mathcal{N}_s}} {\sum\limits_{k \in \mathcal{K}}} \bigg[{{u_{r,k,{n_s}}}(\mathbf{p}(l-1))-{v_{r,k,{n_s}}}(\mathbf{p}(l-1))} \\
\begin{aligned}
& -\nabla{{v_{r,k,{n_s}}}(\mathbf{p}(l-1))}(\mathbf{p}(l-1)-\mathbf{p}(l-1))\bigg]\\
\end{aligned}
\end{multline*}
	\begin{equation*}
		= \sum\limits_{r \in \mathcal{R}} {\sum\limits_{s \in \mathcal{S}} {\sum\limits_{{n_s} \in \mathcal{N}_s} {\sum\limits_{k \in \mathcal{K}} {{u_{r,k,{n_s}}}(\mathbf{p}(l-1))-{v_{r,k,{n_s}}}(\mathbf{p}(l-1))}}}}.
	\end{equation*}
	Thus,  after each iteration $l$ of the proposed SCA approach, the objective value of (\ref{Power alloction problem}) is improved (increased) or remains unchanged after iteration $l$. As a result, the SCA approach will converge to a locally optimal solution $\mathbf{p^*}$.

\section*{References}


\end{nolinenumbers}
\end{document}